# Autonomic Vehicular Networks: Safety, Privacy, Cybersecurity and Societal Issues

Gerard Le Lann
*INRIA*
Paris-Rocquencourt, France
gerard.le_lann@inria.fr

*Abstract*—Safety, efficiency, privacy, and cybersecurity can be achieved jointly in self-organizing networks of communicating vehicles of various automated driving levels. The underlying approach, solutions and novel results are briefly exposed. We explain why we are faced with a crucial choice regarding motorized society and cyber surveillance.

I. INTRODUCTION

Vehicular networks are bound to be autonomic, i.e. self-organizing and self-healing. An autonomic vehicular network (AVN) is comprised of partially or fully automated vehicles, assigned SAE automated driving levels ranging from 0 to 5 [1]. The focus of this paper is on AVN constructs and vehicular capabilities meant to achieve the following four fundamental properties: *safety* (a significant reduction of severe injuries, fatalities, and damages), *efficiency* (high vehicular densities, good asphalt utilization ratios), *privacy* (no cyberespionage), and *cybersecurity* (immunity to cyberthreats). With current and upcoming standards such as, e.g., IEEE 802.11p, ETSI ITS-G5, IEEE 1609-2 (MAC and channel management), SAE J2945/1-2.2, ETSI TS 102 637 (periodic beaconing), and ETSI TS 102 940 (security schemes based on PKIs (Public Key Infrastructures)), none of the aforementioned goals can be reached fully. Authors who have arrived at such conclusions have published proposals aimed at improving current solutions. However, intractable limitations remain. For example, current thinking is that one cannot have safety *and* privacy together. That is incorrect. Our intent here is to share a vision of how to meet the challenge of achieving *safety*, *efficiency*, *privacy* and *cybersecurity* jointly, by design. This vision may be considered unreasonably radical by a fraction of the ITS community. However, there is no choice: under current approaches, risks of cyber-surveillance and collisions due to cyberattacks are not eliminated fully. Along with other scientists and lawyers, we advocate for the adoption of novel approaches and solutions that would serve as foundations for the next "wave" of standards, denoted WAVE 2.0 for convenience, WAVE 1.0 referring to current and upcoming standards.

We have been here before. Remember the MAP initiative by General Motors, the IBM Token Ring saga, and other similar failures? Despite huge investments and financial losses, products based on the IEEE 802.4, 802.5, 802.6 (DQDB) standards, or FDDI networks, have disappeared soon after deployment. Whether this could be the fate of 802.11p is not our concern here. Rather, we anticipate the continuous emergence of novel communication technologies as well as the adoption of more rigorous approaches regarding the design of onboard systems and solutions for next-gen AVNs.

Some recent results are presented in this paper. Due to space restrictions, they cannot be exposed thoroughly. Detailed presentations are to appear in forthcoming publications and research reports. We introduce *partitioned architectures for onboard systems* (Section II), *vehicular cells* (Section III), the *trusted vehicular subnetwork construct*, the *spontaneous formation of homogeneous subnetworks of identical SAE automation levels* (Section IV), *PKI-free pseudonym schemes*, *privacy and safety in the presence of cyberthreats*, *physical exclusion of a malicious vehicle*, and the *cyber-stealth mode* (Section V). Adopting or not adopting the WAVE 2.0 approach implies a fundamental societal choice of deep consequences regarding cyber-surveillance.

II. PARTITIONED ONBOARD SYSTEM ARCHITECTURES

AVNs are life-critical systems-of-systems. Conformance to design principles in force in every safety-critical (SC) domain (e.g., aerospace, chemical/nuclear plants) is thus mandatory, notably a strict separation of SC and non-SC functionalities. Fig. 1 shows a simple example of a partitioned onboard system architecture. A SC subsystem comprises a SCR subsystem (robotics and other capabilities, e.g., AI and algorithmic learning), and a SCC subsystem (short-range neighbor-to-neighbor (N2N) and SC V2V radio communications, optical communications, time-bounded coordination algorithms). SC subsystems are equipped with specific processors and real-time operating systems, and they shall be ASIL-D compliant (ISO 26262 standard). Besides its own inputs (radars, lidars, cameras), a SCR subsystem processes data written by a SCC subsystem in their shared memory. They share a UTC-aligned clock (GNSS inputs) and back-up clocks (to cope with GNSS outages) as well as a tamper-proof device (TPD) which records significant events, and may fire a Stop maneuver, to be performed by the SCR subsystem. A non-SC (NSC) subsystem hosts non safety-critical V2X (V2V/V2I) telecommunication capabilities (WAVE 1.0 standards, 4G LTE, 5G, 6G), most of them resting on terrestrial nodes for, e.g., infotainment, access to Internet, cloud services and to PKIs. A NSC subsystem cannot write in the SC subsystem data space. The NSC TPD can be updated via V2X telecommunications.

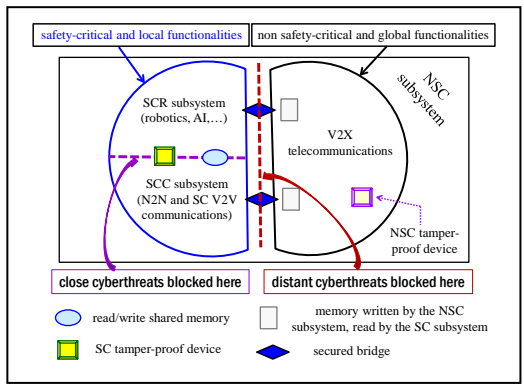

Fig. 1. A partitioned onboard system

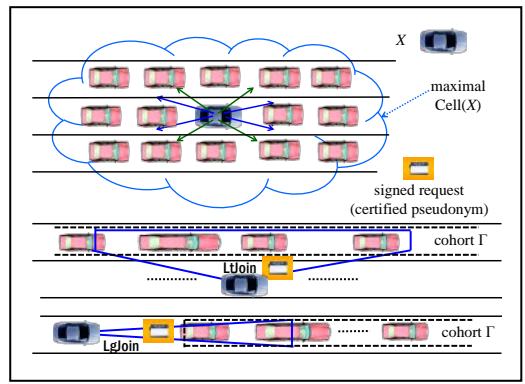

Fig. 2. Vehicular Cell and authenticated Join maneuvers

Terrestrial topology provided in emaps shows lanes and their numbering, starting from 1 for the rightmost lane (leftmost in certain countries). Numbers are assigned to roads at intersections and roundabouts. Knowledge of correct lane-level positioning is achieved by SCR and SCC subsystems, e.g., via hybrid measurements [2] or visual SLAM.

## III. SAFETY AND EFFICIENCY

### A. Vehicular Cells and Safety-Critical Communications

WAVE 1.0 radio ranges in the order of 300 m and reliance on V2I communications are inappropriate for safety. The *vehicular cell* concept derives from observing that vehicles may collide only if close to each other. Let Cell($X$) stand for $X$'s vehicular cell. Assuming a multilane setting, highest traffic density and smallest vehicles size, maximal Cell($X$) would comprise 14 members in addition to $X$ (2 closest predecessors and successors and 5 neighbors in each lateral lane), as shown in Fig. 2. Neighbors that may hit $X$ belong to Cell($X$). In a Cell, collision prevention implies ultra-fast coordination. That is the purpose of N2N very short-range communications (up to 30 m approximately), activated only when needed (no beaconing), longitudinally (e.g., steep deceleration) or laterally (e.g., lane change, subject to distributed agreement). Anonymity is a fundamental feature of N2N messages: source and destination fields carry non-reversible names (no MAC or IP address)— see Section IV. Vehicles that belong to different Cells shall be able to coordinate their behaviors. That is the purpose of short-range SC V2V communications (up to 60 m approximately). For example, safe *entrance* to un-signaled intersections or roundabouts can be achieved by exchanging SC V2V messages among members of competing Cells. (Safe *crossings* would mostly rest on optical communications.) N2N and SC V2V messaging is performed via power-controlled omnidirectional, directional, or MIMO radio antennas that operate on channels distinct from those used for V2X telecommunications. With power control, radio ranges can be tuned as desired (up to ≈ 30 m or ≈ 60 m). N2N and SC V2V communications serve to cope with failures of onboard sensors and to achieve *proactive safety*—the ability to influence the behaviors of nearby vehicles. Communications are not just yet another "sensing" technology. They supplement robotics and AI, able to achieve *reactive safety* only, since behavioral decisions are kept untold—others have to guess. Optics-based N2N messaging is an invaluable technology regarding privacy.

V2X messages that are authenticated can be imported in a SCC subsystem for further processing (otherwise, a V2X message is kept inside a NSC subsystem). Privacy and cybersecurity issues related to V2X telecommunications differ from those arising with N2N and SC V2V communications.

### B. Restricted Spanning of N2N Communications

Technologies change at a higher pace than requirements (our four properties). It is thus wiser to specify desired communication primitives, leaving open the choice of appropriate underlying technologies. This is consistent with the emergence of software-defined architectures and with our recommendation for top-down approaches. Four primitives are defined for N2N messaging. A LgSend primitive serves to trigger range 2 longitudinal communications: a N2N message sent by $X$ shall reach $X$'s 2 closest predecessors and successors, if any. A LtSend primitive serves to trigger range 1 lateral communications: $X$ shall be able to send N2N messages to its closest adjacent (left, right) neighbors, if any. Communications are transitive across Cells: N2N messages can be disseminated longitudinally and laterally. Symmetrically, we have defined the LgReceive and LtReceive primitives (see Section IV). Enforcing restricted spanning for N2N communications has decisive merits regarding cybersecurity.

### C. MAC Protocols

A MAC protocol shall match higher-level requirements. As for safety and efficiency, a MAC protocol shall guarantee small *bounded* channel access delays ($\lambda$) under *highest channel contention conditions* and small bounded delays for message dissemination under highest message loss conditions. A MAC protocol is acceptable only if distances travelled in $\lambda$ are an order of magnitude smaller than inter-vehicular gaps. As for privacy and cybersecurity, a MAC protocol shall help in *thwarting eavesdropping, masquerading, message forging, Sybil attacks*, and the like. WAVE 1.0 MAC protocols fail to meet these requirements, by huge margins concerning $\lambda$ [3-6]. Different MAC protocols are defined for N2N radio channels and for SC V2V radio channels. Suitable protocols can be found in the literature—e.g., [7-8].

### D. Periodic Beaconing Considered Useless for Safety

Given that radio communications are unreliable, some vehicle(s) may not receive beacons delivered to others. Consequently, Local Dynamic Maps (LDMs) built out of periodic beaconing (Cooperative Awareness Messages) may differ significantly, which is a serious weakness regarding safety. Consider *X*, *A* and *B* moving at 90 km/h. *X* broadcasts beacons at 1 Hz. Consider the case where a given beacon received by *A* is not received by *B*. *X*'s geolocations recorded in LDM(*A*) and LDM(*B*) differ by 25 m, enough for creating collisions. Despite periodic updates, discrepancies may persist. Algorithms such as terminating atomic broadcast or consensus keep LDMs mutually consistent (identical in our case), under assumptions … that are not met in AVNs. In an asynchronous model (delays have unknown bounds—the case with WAVE), consensus is impossible in the presence of a single loss [9]. In a synchronous model (delays have known upper bounds), neither unanimity nor strong majority can be achieved when loss ratios exceed modest thresholds [10-11]. Since channel access delays differ for any 2 broadcasters, geolocations read in different beacons cannot mirror a correct instantaneous "picture" of some real global state. (This is a problem studied over the last 30 years in the Distributed Algorithms community.) Being possibly inaccurate and mutually inconsistent, LDMs are useless for making safety-preserving decisions. Moreover, LDMs cannot help in avoiding accidents, since vehicle trajectories cannot be predicted with certainty from past trajectories. A LDM updated at UTC time *t* cannot reflect moves made by vehicles right after *t*, which moves may be dictated by unexpected hazards—besides depending on each other. Finally, why should a vehicle waste communication and computational resources to maintain knowledge relative to distant vehicles that it will never see? Periodic beaconing is just another passive solution that may help to achieve reactive safety probabilistically, when proactive safety is needed for collision prevention. In addition to being useless (and unreliable), periodic beaconing may be harmful. Beacons carry unencrypted GNSS data (in particular) that can be heard by distant unknown vehicles and terrestrial nodes, thus facilitating eavesdropping and tracking.

## IV. COHORTS

Efficiency (very small inter-vehicular gaps) is antagonistic with safety. Both properties can nevertheless be reconciled. An AVN being an instance of a cyber-physical system, correct solutions for safety and efficiency must be based on some appropriate cyber-physical construct. A linear formation of vehicles in a given lane is commonly referred to as a platoon (long-lived formation) or as a string (short-lived ad hoc formation). In [12-13], we have presented and discussed the cohort construct, which is a linear formation endowed with a specification. In addition to small inter-vehicular gaps (iv-gaps), safe inter-cohort gaps (IC-gaps) have been proposed, large enough for guaranteeing that the head of a cohort never hits the tail of a preceding cohort. The rationale is to keep the number of rear-end collisions below an acceptable threshold in case a "brick wall" condition would develop within a cohort. A spontaneous ranking of members is a crucial property of the cohort construct. An example of a similar idea can be found in [14]—concepts of convoy, authenticated members and vehicle sequence numbers. In a cohort, members assign themselves consecutive ranks, 1 for the head and *n* (an inverse function of velocity) for the tail. Highest velocities and cohort sizes depend on the setting (city streets, country roads, highways). Any 2 contiguous members exchange N2N messages that carry SC data or data needed for internal cohort management (e.g., common knowledge of current *n* and velocity). Time-bounded broadcasts can be instantiated via cohort-wide N2N message dissemination [15]—unfeasible with WAVE 1.0 solutions. Details relative to reliability (losses of N2N messages, cohort splits performed when a N2N link is detected failed) can be found in referenced publications. Names that appear in N2N messages exchanged within a cohort circulating in lane numbered *j* are pairs of integers {*r*,*j*}, where *r* stands for sender's rank. Longitudinal N2N messages processed by a member ranked $r^{th}$ are those received from neighbors of ranks *r*+1 and *r*+2, or/and *r*-1 and *r*-2 (the LgReceive primitive). As for lateral N2N messaging and LtReceive, designation of specific lateral neighbors without knowing their names is an issue. This problem has original solutions, based on optics [16-17]. Human-less seeing-is-believing [17] fits ideally with the cohort construct (to appear). A vehicle that wants to become member of a cohort activates a LgJoin primitive (if catching up with the tail of a cohort) or a LtJoin primitive (in case of a lane change), based on LgSend and LtSend primitives (see Fig. 2).

A novel result relates to the spontaneous formation of homogeneous cohorts. Let *aul*(*V*) stand for *V*'s automated driving level. Assume that vehicles reactivity and agility are higher with higher automation levels. With smaller iv-gaps and IC-gaps, efficiency is higher as well. A cohort ($\Gamma$) specification stipulates the smallest and highest automation levels accepted in $\Gamma$, denoted *SL*($\Gamma$) and *HL*($\Gamma$), respectively. (These levels may change dynamically, subject to cohort-wide agreement.) LgJoin and LtJoin primitives have *aul*(.) as a parameter. If $SL(\Gamma) \leq aul(V) \leq HL(\Gamma)$ holds, then *V* is accepted in cohort $\Gamma$. A strictly homogeneous cohort $\Gamma$ is spontaneously created by setting $SL(\Gamma) = HL(\Gamma)$. There is no need for restricting the circulation of highly automated vehicles to specific lanes. In addition to being antagonistic with the efficiency property, such a constraint might be a source of frustration. In a multilane highway, other vehicles may enjoy fluid traffic conditions while congestion would develop in specific lanes. Vehicles assigned low automation levels may or may not be equipped with communication capabilities. In the former case, they would form inefficient cohorts (large iv-gaps). Inefficiency is highest in the latter case (human-driven vehicles), since we would have iv-gaps = IC-gaps. Another property of cohorts has been uncovered recently: a cohort is a trusted vehicular sub-network.

## V. PRIVACY AND CYBERSECURITY

Originally devised for *safety* and *efficiency*, a cohort is essential also for *privacy* and *cybersecurity*. Note that issues of cyber-surveillance carried out by sensors installed inside vehicles (e.g., cameras) are out of the scope of this paper.

### A. Pseudonymity

Solutions based on asymmetric cryptography, PKIs and pseudonymity [18] are being worked out by standard-making bodies. When registered and authenticated by a certification authority (CA), a vehicle is assigned a unique reversible certificate for accountability purposes (auditability, liability) as well as credentials, i.e. pairs {pseudonym, pseudonym certificate}, referred to as a (certified) pseudo here. Current solutions are not satisfactory [19-21]. For example, tracking is feasible despite frequent changes of pseudos. Major roadblocks can be eliminated as follows: no binding with beaconing, no dependence on PKIs (no importation of new pseudos). Why should a vehicle "prove" that it has been authenticated each time it sends a message? That is unnecessary if, as commonly assumed, most vehicles are not malicious. With vehicular networks that lack structuring constructs and that rely on V2X telecommunications, there might be no alternative. Pseudos are spent at high frequencies, and vehicles must replenish their TPDs with new credentials. As a result, an AVN is prone to man-in-the-middle (MitM) cyberattacks (e.g., suppression or falsification of pseudos) launched via intrusions of terrestrial nodes. Moreover, communication and computational resources are wasted. Consider ECDSA-256 signatures, a verification time of 2 ms, 7 Hz beaconing frequency, and a vehicle within WAVE 1.0 radio range ($\approx$ 300 m) of $x$ vehicles. Thrashing develops as $x$ tends to 70, the case in numerous settings. There are better ways of using onboard resources. Finally, assuming that vehicles are always within reach of terrestrial nodes, that they enter mix-zones whenever necessary, and so on severely restricts the applicability of current solutions.

*B. Anonymity in Trusted Vehicular Subnetworks*

A cohort is a vehicular subnetwork where members can trust each other, provided that admission into a cohort is subject to verified authentication, as shown in Fig. 2. A pseudo must be used to sign a LgJoin or a LtJoin request. Owing to controlled admission, no pseudo need be used afterwards for N2N messaging. As a result, pseudos available in a SC-TPD are consumed at reasonable a rate. This removes the need for on-the-move TPD replenishment of new pseudos via PKIs. If required by administrative or contractual regulations, TPD replenishment would be performed at secured places (no V2I communications). Vehicle thefts (an issue with long-lived pseudos) can be prevented or combated very easily. Self-issued names $\{r,j\}$ that appear in N2N messages are non-reversible. Full anonymity is thus guaranteed, and these names entail very negligible processing overhead. WAVE 2.0 solutions provide vehicles with a double obfuscation scheme (pseudos and names $\{r,j\}$), while meeting the accountability requirement. Another condition for intra-cohort trust is immediate detection of any malicious behavior (see Subsection E).

*C. Close Cyberthreats in Trusted Vehicular Subnetworks*

Owing to the LgReceive and LtReceive primitives (Section IV), close cyberthreats that target $X$ may only be launched by members of Cell($X$). That is a drastic change compared to traditional approaches where cyberthreats may be instigated by unknown distant vehicles and terrestrial nodes. Close eavesdropping is not an issue. N2N messages do not carry GNSS coordinates, thus geo-privacy is ensured. N2N messages carry unencrypted SC data such as "lane merging", "clear lane" (emergency vehicle), "new $v$" (bad road conditions, lane blocking). Secrecy is not an issue. Vehicles nearby Cell($X$) may overhear N2N messages exchanged by members of Cell($X$) which start a SC maneuver. That can only be beneficial (think of concurrent conflicting maneuvers). As regards metadata, close tracking of a given name $\{r,j\}$ is pointless, since the same name may be used by different vehicles at unpredictable different times. Assuming in the order of 5,000 pseudos in a SC-TPD, none would be used more than once a day. Reuse is thus possible without being tracked. Close *physical* tracking/spying (a vehicle is followed by the same car for too long) is out of the scope of this paper.

We have analyzed close cyberattacks such as, e.g., N2N message falsification, suppression, masquerading, Sybil attack, injection of false messages, and demonstrated that they cannot endanger safety in a Cell or defeat controlled admission in a cohort (to appear). Consequently, they can only originate from irrational neighbors. A honest vehicle may behave as an irrational adversary when its SCC subsystem has been contaminated by a malware, leading to those mitm (malware-in-the-middle) cyberattacks mentioned above. A contaminated vehicle is quickly excluded (see Subsection E). It is customary to assume that an AVN comprises a minority of malicious vehicles. Assuming cohorts of size 3 at least (other cases not detailed here, for conciseness), this assumption can be stated more accurately: at most 1 malicious member in any group of 3 longitudinal neighbors. Owing to the LgReceive primitive, forging or suppression of a longitudinal N2N message $m$ sent by a range 2 neighbor and relayed by a range 1 neighbor is trivially detected: $m$ is accepted only if seen twice. A lateral N2N message is accepted only if identical or consistent copies are received from at least 2 out of 3 neighbors. Masquerading and Sybil attacks are detected on the spot, owing to a generic property of suitable MAC protocols: $X$ of rank $x$ can only send a longitudinal (resp., lateral) N2N message at a specific UTC time $Lg_x$ (resp., $Lt_x$). A receiver can thus trivially tell whether $X$ tries to impersonate another member (radio channel accessed at times other than $Lg_x$ or $Lt_x$, or rank other than $x$). The same goes for injections of false messages.

*D. Distant Cyberthreats and Cyber-Stealth Mode*

They can be launched via V2X communications. Given that a NSC subsystem has no access to a SCR subsystem and that motions of a vehicle are under the exclusive control of a SCR subsystem, distant cyberattacks cannot compromise safety. Recall that SC-TPDs are immune to distant cyberattacks (no new pseudos need be imported). Conversely, cybersecurity is not guaranteed for NSC subsystems. Distant cyberattacks on input ports may disrupt convenient services (e.g., infotainment, traffic data), a tolerable nuisance. Distant eavesdroppers may exploit outgoing V2X messages. Personal data secrecy/privacy (message contents) can be achieved through encryption. Other privacy threats are linked with metadata. Tracking is feasible despite frequent pseudo changes, and facilitated with probes issued by commodity-grade wifi handlers. We introduce the cyber-stealth mode option for NSC subsystems of privately owned vehicles: muteness is enforced (no V2X outputs, to the exception of e-Call and pseudonymized crowdsourcing—see further). This option which guarantees passengers privacy is in line with the EU General Data Protection Regulation entered in force in Europe in May 2018 [22]. SC V2V messages serve to *avoid* accidents. Thus, medium-range V2V/X2V emergency messages broadcast for announcing an accident (*recognition of a defeat*) are not SC messages. Such messages may not be necessary since recent radars and cameras have ranges comparable to WAVE ranges. The broadcasting of V2X messages is useful for updating global traffic data. *Estimates* suffice for planning and traffic management purposes. A variation of periodic beaconing akin to crowdsourcing may be of interest. Every member of a cohort (Γ) runs an algorithm *A* serving to tell whether it must broadcast a V2X message carrying Γ's length (common knowledge within Γ) and its GNSS coordinates. A pseudo found in its NSC-TPD is used to sign such a message. Receivers learn about the existence of a cohort, its spanning and approximate geolocations of non-identifiable members. Besides preserving privacy, this scheme induces light channel loads: 1 (deterministic *A*) or much less than *n* (probabilistic *A*) broadcasts at every algorithmic round.

*E. Malicious Behavior, Physical Exclusion, Accountability*

That a CA or a PKI has delivered pseudos to a vehicle *X* does not prohibit *X* from behaving maliciously, in the physical space, and in cyber space by injecting bogus data, thus causing accidents. Revocation of credentials (PKI-based solutions) achieves eviction in cyber space, which does not suffice. A malicious vehicle must be *physically* excluded from an AVN. To this end, SC-TPDs contain predicates, such as:

- Any 2 consecutive activations of LgSend or LtSend shall be separated at least by (latency derived from a MAC protocol)

- Member ranked $r^{th}$ may activate LgSend (resp., LtSend) at UTC times $Lg_x$ (resp., $Lt_x$) only

- SCR subsystem sensors detect a new neighbor in a Cell and N2N communications are active.

Violation of a predicate, UTC timestamped, is recorded in SC-TPD, and leads to a Stop. The SCR subsystem, informed by the SCC subsystem, takes control: red flashing warning lights are turned on, velocity is reduced, until a safe halt is possible (parking spot or emergency lane). A pseudonymized and encrypted V2X message is broadcast, aimed at appropriate public/private organizations (e.g., police, highway authorities, insurance companies). That message carries the GNSS location of where to find the vehicle, its reversible certificate, and other data recorded in SC-TPD. Accountability is guaranteed.

*F. Societal Issues*

With WAVE 1.0 solutions, safety can be compromised by cyberattacks, and cyber-surveillance remains an issue. Fees are charged for telecommunications and PKI services. WAVE 2.0 solutions—those briefly exposed here or found in recent publications, and under study—achieve highest safety and immunity to cyberattacks. Full privacy is enabled with the cyber-stealth mode option. Modest fees are due for V2X telecommunications, and there is no billing for N2N or SC V2V communications or PKI services. The crucial question is:

In which type of motorized society do we want to live, subject to or free from cyber-surveillance?